\documentclass[article,column]{IEEEtran}

\usepackage{amsmath, amssymb, setspace}

\newcommand{\y}{\mathbf{y}}
\newcommand{\condexp}{E(\theta|\y)}
\newcommand{\deltanaught}{\delta_0(\y)}
\newcommand{\psif}{\psi(\y,\theta)}
\newcommand{\psifsq}{\psi^2(\y,\theta)}
\newcommand{\btheta}{\boldsymbol{\theta}}
\newcommand{\g}{\mathbf{g}}
\newcommand{\bdelta}{\boldsymbol{\delta}}
\newcommand{\covcond}{\boldsymbol{\Sigma}_{\y}}
\newcommand{\cov}{\boldsymbol{\Sigma}}
\newcommand{\bpsi}{\boldsymbol{\psi}}
\newcommand{\bepsilonzero}{\boldsymbol{\epsilon}_0}
\newcommand{\error}{ \theta - \deltanaught }

\newcommand{\hinv}{ \frac{1}{h} }
\newcommand{\score}{ \frac{\partial \log p(\y|\theta)}{\partial \theta} }
\newcommand{\dpdtheta}{ \frac{\partial p(\y|\theta)}{\partial \theta} }
\newcommand{\var}{\text{var}}

\ifCLASSINFOpdf
\else
\fi
\begin{document}
%
\title{Inequalities for the Bayes Risk}
%
%
%

\author{
~Asaf~Weinstein
and
Ehud~Weinstein,~\IEEEmembership{Fellow,~IEEE}

\thanks{A. Weinstein is with the Department of Statistics, the Wharton School of the University of Pennsylvania, Philadelphia, PA 19104, U.S.A (asafw@wharton.upenn.edu). 
E. Weinstein is with the School of Electrical Engineering, Tel-Aviv University, Tel-Aviv 69978, Israel (udi@tau.eng.ac.il).}

}

\maketitle

\begin{abstract}
Several inequalities are presented which, in part, generalize inequalities by Weinstein and Weiss, giving rise to new lower bounds for the Bayes risk under squared error loss.
\end{abstract}


%
\IEEEpeerreviewmaketitle

%
%
%
%

Consider a Bayesian setting in which $\y$ denotes the observed data and $\theta$ denotes an unknown scalar parameter.
Suppose that $E\left[ (\theta - \condexp)^2 | \y \right]$ and $E\left[ (\theta - \condexp)^2 \right]$ exist.
For any estimator $\delta(\y)$ of $\theta$, 
\begin{align}
E\left[ (\theta - \delta(\y))^2 |\y \right] &\geq E\left[ (\theta - \condexp)^2 | \y \right] \label{eq:bayes cond} \\
E\left[ (\theta - \delta(\y))^2 \right] &\geq E\left[ (\theta - \condexp)^2 \right] \label{eq:bayes}
\end{align}
The right hand sides of \eqref{eq:bayes cond} and \eqref{eq:bayes} are the ultimate lower bounds as they are attainable. 
However, these bounds are often difficult to analyze or even to compute. 
Consequently, numerous lower bounds have been proposed that are presumably simpler to analyze and compute. 
A comprehensive survey of various approaches can be found in~\cite{van2007bayesian}. 

Starting off with the approach in~\cite{ww85a}, let $\psif$ be any appropriately measurable function of $\y$ and $\theta$. 
Invoking Cauchy-Schwarz inequality, 
\begin{equation} \label{eq:cs}
E\left[ (\theta - \condexp)^2 \right] \geq \frac{ E^2 \left[ (\theta - \condexp) \psif \right ] }{\var \left[ \psif \right]}
\end{equation}
provided that $\var \left[ \psif \right]$ exists and is strictly positive. 
Combining \eqref{eq:cs} with \eqref{eq:bayes},
\begin{equation} \label{eq:LB}
E\left[ (\theta - \delta(\y))^2 \right] \geq \frac{ E^2 \left[ (\theta - \condexp) \psif \right ] }{\var \left[ \psif \right]}
\end{equation}

Simliarly,
\begin{equation} \label{eq:cs cond}
E\left[ (\theta - \condexp)^2 |\y \right] \geq \frac{ E^2 \left[ (\theta - \condexp) \psif \right | \y] }{\var \left[ \psif | \y \right]}
\end{equation}
provided that $\var \left[ \psif | \y \right]$ exists and is strictly positive.
Hence, 
\begin{equation} \label{eq:cs avg cond}
E\left[ (\theta - \condexp)^2 \right] \geq E\left\{  \frac{ E^2 \left[ (\theta - \condexp) \psif \right | \y] }{\var \left[ \psif | \y \right]}  \right\}
\end{equation}
Combining \eqref{eq:cs cond} with \eqref{eq:bayes cond},
\begin{equation} \label{eq:LB cond}
E\left[ (\theta - \delta(\y))^2 |\y \right] \geq \frac{ E^2 \left[ (\theta - \condexp) \psif \right | \y] }{\var \left[ \psif | \y \right]}
\end{equation}
Combining \eqref{eq:cs avg cond} with \eqref{eq:bayes},
\begin{equation} \label{eq:LB avg cond}
E\left[ (\theta - \delta(\y))^2 \right] \geq E\left\{  \frac{ E^2 \left[ (\theta - \condexp) \psif \right | \y] }{\var \left[ \psif | \y \right]}  \right\}
\end{equation}

At last, 
%
\begin{equation} \label{eq:cs cond theta}
E\left[ (\theta - \condexp)^2 |\theta \right] \geq \frac{ E^2 \left[ (\theta - \condexp) \psif \right | \theta] }{E \left[ \psifsq | \theta \right]}
\end{equation}
provided that $E\left[ (\theta - \condexp)^2 |\theta \right]$ exists and that $E \left[ \psifsq | \theta \right]$ exists and is strictly positive.
Hence,
\begin{equation} \label{eq:cs avg cond theta}
E\left[ (\theta - \condexp)^2 \right] \geq E \left\{    \frac{ E^2 \left[ (\theta - \condexp) \psif \right | \theta] }{E \left[ \psifsq | \theta \right]}    \right\}
\end{equation}
Thus, combining \eqref{eq:cs avg cond theta} with \eqref{eq:bayes},
\begin{equation} \label{eq:LB avg cond theta}
E\left[ (\theta - \delta(\y))^2 \right] \geq E \left\{    \frac{ E^2 \left[ (\theta - \condexp) \psif \right | \theta] }{E \left[ \psifsq | \theta \right]}    \right\}
\end{equation}

The extension of inequalities \eqref{eq:LB}, \eqref{eq:LB cond}, \eqref{eq:LB avg cond} and \eqref{eq:LB avg cond theta} to multiple parameters and any functions of the parameters is given in Appendix~\ref{app:generalization}.

\bigskip

If 
\begin{equation} \label{eq:condition}
E\left[ \psif |\y \right] = 0
\end{equation}
then \eqref{eq:LB} becomes
\begin{equation} \label{eq:WWI}
E\left[ (\theta - \delta(\y))^2 \right] \geq \frac{ E^2 \left[ \theta \psif \right] }{E \left[ \psifsq \right]}
\end{equation}
and \eqref{eq:LB cond} becomes
\begin{equation} \label{eq:WWI cond}
E\left[ (\theta - \delta(\y))^2 |\y \right] \geq \frac{ E^2 \left[ \theta \psif \right | \y] }{E \left[ \psifsq | \y \right]}
\end{equation}
which are the inequalities presented in~\cite{ww85a} and~\cite{weissphd}. 
Hence, \eqref{eq:WWI} and \eqref{eq:WWI cond} are special cases of \eqref{eq:LB} and \eqref{eq:LB cond}, respectively, obtained by restricting $\psi$ to satisfy \eqref{eq:condition}.

The tightest lower bounds are obtained by maximizing the right hand side of the inequalities above with respect to $\psi$. The solution in all cases is
\begin{equation} \label{eq:trivial psi}
\psi(\y,\theta) = \theta - \condexp
\end{equation}
in which case \eqref{eq:LB}, \eqref{eq:LB avg cond} and \eqref{eq:LB avg cond theta} reduce to \eqref{eq:bayes}, and \eqref{eq:LB cond} reduces to \eqref{eq:bayes cond}. 
Our objective is to find instead functions $\psi$ that may produce somewhat weaker lower bounds but, in turn, may be simpler to analyze and compute.

Weiss and Weinstein (~\cite{weissphd}~\cite{ww85a}) suggested to use
\begin{equation} \label{eq:psiWW}
\psi( \y, \theta )= 
\begin{cases}
\left( \frac{p(\y, \theta+h)}{p(\y, \theta)} \right)^s - \left( \frac{p(\y, \theta-h)}{p(\y, \theta)} \right)^{1-s} &p(\y, \theta)>0 \\
0 & \text{o.w.}
\end{cases}
\end{equation}
where $p(\y, \theta)$ denotes the joint probability density or probability mass of $\y$ and $\theta$. It can be verified that this choice of $\psi$ satisfies the condition in \eqref{eq:condition} for any combination of $h$ and  $0<s<1$. 
Substituting \eqref{eq:psiWW} into \eqref{eq:WWI}, we obtain the so-called Weiss-Weinstein family of lower bounds (~\cite{ww85a}~\cite{weissphd}~\cite{ww85b}~\cite{ww88}), indexed by the variables $h$ and $s$. As was demonstrated in a variety of examples and applications (see~\cite{van2007bayesian} and references therein), the Weiss-Weinstein bound is relatively easy to analyze and compute, while optimizing with respect to $s$ and $h$ may produce tight lower bounds.

As an alternative to \eqref{eq:psiWW} consider
\begin{equation} \label{eq:psinew}
\psi( \y, \theta )=
\begin{cases}
\left( \frac{p(\y | \theta+h)}{p(\y | \theta)} \right)^s - \left( \frac{p(\y | \theta-h)}{p(\y | \theta)} \right)^{1-s} &p(\y | \theta)>0 \\
0 & \text{o.w.}
\end{cases}
\end{equation}
where $p(\y | \theta)$ denotes the conditional probability density or probability mass of $\y$ given $\theta$. This choice of $\psi$ does not satisfy the condition \eqref{eq:condition} and, hence, inequalities \eqref{eq:WWI} and \eqref{eq:WWI cond} cannot be used. However, all other inequalities still apply. 
Specifically, using \eqref{eq:psinew} in \eqref{eq:LB avg cond theta} we obtain a new family of lower bounds indexed by the variables $h$ and $s$. 
As sketched in Appendix~\ref{app:asymptotic optimality proof}, asymptotically, under certain regularity conditions, this family of lower bounds converges in the limit as $h \rightarrow 0$ to
\begin{equation} \label{eq:asymptotic optimality}
E\left[ (\theta - \delta(\y))^2 \right] \geq E \left\{    \frac{ 1 }{E \left[ (\score)^2 | \theta \right]}    \right\}
\end{equation}
This bound is asymptotically attainable, a property that has not been proven to hold with respect to the Weiss-Weinstein family. Of course, if the asymptotic conditions are not satisfied the bound in \eqref{eq:asymptotic optimality} may be a weak lower bound, and if the regularity conditions are not satisfied this bound may not even exist. In such circumstances, optimizing with respect to $s$ and $h$, it may be possible to obtain a significantly tighter lower bound, free of the associated regularity conditions.

\appendices
\numberwithin{equation}{section}

\section{} \label{app:generalization}

Define
\begin{align*}
&\cov \triangleq E\left[ (\g-\bdelta) (\g-\bdelta)^T\right] \\
&\covcond \triangleq E\left[ (\g-\bdelta) (\g-\bdelta)^T | \y \right] \\
&\bepsilonzero \triangleq \g - E(\g|\y).
\end{align*}
where $\btheta$, $\g=\g(\btheta)$, $\delta = \delta(\y)$, and $\bpsi = \bpsi( \y, \btheta )$ are vector quantities.
Applying the vector form of Cauchy-Schwarz inequality and following the same considerations leading to \eqref{eq:LB}, \eqref{eq:LB cond}, \eqref{eq:LB avg cond} and \eqref{eq:LB avg cond theta}, under analogous conditions we obtain, respectively,
\begin{align}
&\cov \geq E [ \bepsilonzero \bpsi ^T ] \ Cov^{-1} [\bpsi ] \ E [ \bpsi \bepsilonzero^T ] \label{eq:three} \\
&\covcond \geq E [ \bepsilonzero \bpsi ^T | \y ] \ Cov^{-1} [\bpsi | \y ] \ E [ \bpsi \bepsilonzero^T |\y ] \\
&\cov \geq E \{ E [ \bepsilonzero \bpsi ^T | \y ] \ Cov^{-1} [\bpsi | \y ] \ E [ \bpsi \bepsilonzero^T |\y ] \} \\
&\cov \geq E \{ E [ \bepsilonzero \bpsi ^T | \btheta ] \ Cov^{-1} [\bpsi | \btheta ] \ E [ \bpsi \bepsilonzero^T |\btheta ] \}
\end{align}
Note that the dimensions of $\btheta$, $\g$ and $\bpsi$ need not be the same, e.g., $\btheta$ may still be a scalar quantity whereas $\bpsi$ may be a vector.

\section{} \label{app:asymptotic optimality proof}
Let $\psi(\y,\theta)$ be the function defined in \eqref{eq:psinew}. 
To simplify the exposition let $s=1$, in which case 
\begin{equation}
\frac{1}{h} \psif = 
\begin{cases}
\frac{1}{p(\y | \theta)} \frac{p(\y | \theta + h) - p(\y | \theta)}{h} & p(\y | \theta)>0 \\
0 & \text{o.w.}
\end{cases}
\end{equation}
If $\psif$ is differentiable with respect to $\theta$ a.e. $\y \in \Omega_{\theta} \triangleq \{ \y: p(\y|\theta) > 0 \}$ then, under the appropriate regularity conditions,
\begin{equation} \label{eq:denom}
\lim_{h\rightarrow 0} E \left[ (\hinv \psif)^2 | \theta \right] = E \left[ (\score)^2 | \theta \right]
\end{equation}
and
\begin{align} \label{eq:num}
& \lim_{h\rightarrow 0} E \left[ (\error) \hinv \psif | \theta \right] \notag \\
&= \theta E \left[ \score | \theta \right] - E \left[ \deltanaught \score | \theta \right] \notag \\
&= \theta \int_{\Omega_{\theta}} \dpdtheta \,d\y - \int_{\Omega_{\theta}} \deltanaught \dpdtheta \,d\y
\end{align}
where $\deltanaught = \condexp$ and where we assumed for simplicity that $p(\y|\theta)$ is a probability density function for every $\theta$. 

If $\Omega_{\theta}$ does not depend on $\theta$,
\begin{equation} \label{eq:termI}
\int_{\Omega_{\theta}} \dpdtheta \,d\y = \frac{\partial}{\partial \theta} \int_{\Omega_{\theta}} p(\y | \theta) \,d\y = 0
\end{equation}
and
\begin{align} \label{eq:termII}
\int_{\Omega_{\theta}} \deltanaught \dpdtheta \,d\y &= \frac{\partial}{\partial \theta} \int_{\Omega_{\theta}} \deltanaught p(\y | \theta) \,d\y \notag \\
& = \frac{\partial}{\partial \theta} E [ \deltanaught | \theta ].
\end{align}
Asymptotically, under appropriate conditions, $E [ \deltanaught | \theta ] \rightarrow \theta$, implying that $\frac{\partial}{\partial \theta} E [ \deltanaught | \theta ] \rightarrow 1$ 
as long as the limit and expectation operations are interchangeable. 
Substituting \eqref{eq:termI} and \eqref{eq:termII} into \eqref{eq:num}, 
\begin{equation} \label{eq:num2}
\lim_{h\rightarrow 0} E \left[ (\error) \hinv \psif | \theta \right] = -1
\end{equation}
Finally, substituting \eqref{eq:denom} and \eqref{eq:num2} into \eqref{eq:LB avg cond theta} and assuming that all limits exist and that $E \left[ (\score)^2 | \theta \right]$ is strictly positive for all $\theta$, we obtain \eqref{eq:asymptotic optimality}.

\ifCLASSOPTIONcaptionsoff
  \newpage
\fi


\bibliographystyle{IEEEtran}
\bibliography{Bibliography}

%
%
%
%
%
%
%

\end{document}